\documentclass[twocolumn, showpacs,letter,aps,a4]{revtex4}
\usepackage{graphicx}
\usepackage{amsmath}

\newcommand{\av}[1]{ {\left\langle #1 \right\rangle} }

\newcommand{\ket}[1]{ {| #1 \rangle} }
\newcommand{\bra}[1]{ {\langle #1 |} }
\newcommand{\rd}{ {\textrm{d}} }

\begin{document}

\title[Noise in an SSET-resonator
driven by an external field]{Noise in an SSET-resonator
driven by an external field}%
\author{D. A. Rodrigues$^{1}$}
\email{denzil.rodrigues@nottingham.ac.uk} \affiliation{$^1$School of
Physics and Astronomy, University of
Nottingham, Nottingham NG7 2RD, U.K.}%
\author{G. J. Milburn$^2$}

\affiliation{$^2$Department of Physics, The University of Queensland, St Lucia, QLD 4072, Australia.}%

\pacs{85.85.+j, 85.35.Gv, 74.78.Na,73.23.Hk}

\begin{abstract}
We investigate the noise properties of a superconducting single
electron transistor (SSET) coupled to an harmonically driven
resonator. Using a Langevin equation approach, we calculate the
frequency spectrum of the SSET charge and calculate its effect on
the resonator field. We find that the heights of the peaks in the
frequency spectra depend sensitively on the amplitude of the
resonator oscillation and hence suggest that the heights of these
peaks could act as a sensitive signal for detecting the small
changes in the amplitude of the drive. The previously known results
for the effective amplitude-dependent damping and temperature
provided by the SSET for the case of a low frequency resonator are
generalized for all resonator frequencies.
\end{abstract}

\maketitle
%


Over the past few years experiments on superconducting circuits have
produced some rather impressive results. Superconducting elements
have been used to produce two-level systems of various kinds which
can be considered as artificial atoms \cite{Nakamura, Martinis,
Cottet, Mooij}, and superconducting stripline resonators can act
analogously to optical cavities for microwave fields \cite{SCR}.
This field of study has been referred to as circuit QED in analogy
with cavity QED. One advantage of superconducting circuits is that
rather than simply investigating the behavior of the system through
the field emitted or reflected by the cavity, other elements can
provide additional information. Mesoscopic conductors coupled to the
resonator can provide additional information and at the same time
the back action may lead to some interesting and subtle dynamics.
Such back-action dynamics have also received considerable attention
in the context of a mesoscopic conductor used to investigate the
behavior of a mechanical resonator \cite{Blencowe,MM,CG,Swede,mmh,
ABZ,SET-expt,ra}.

One system that is of particular interest is that of a
superconducting single electron transistor (SSET) coupled to a
resonator, either mechanical
\cite{bia,SSET1,bc,SSET-expt2,SSET-expt,ria,riha} or composed of a
superconducting stripline \cite{Tsai07}. The coherent transport
through this device at the Josephson quasiparticle resonance
\cite{Choi,CSSET} (JQP) allows a very low-noise current, and at the
same time the sensitivity of the SSET to charge means that the
resonator-SSET coupling is significant. The SSET biased at the JQP
resonance can be considered analogous to a three-level atom, and the
coupled system therefore shows behavior related to that of a
micromaser \cite{micromaser, micromaser_review}.

The low noise and relatively strong coupling in this device allows
the observation of the non-trivial coupled dynamics that arises from
the interaction between the resonator and SSET. When the SSET is
biased so as to absorb energy from the resonator, the SSET acts as
an additional source of damping, and the resonator can be cooled
below its physical temperature. This modification of the steady
state of the resonator due to a back-action induced effective
damping and temperature has been observed for an SSET coupled to
both mechanical and superconducting resonators \cite{SSET-expt2,
Tsai07}. When the SSET is biased to the other side of the JQP
resonance, energy is transferred from the SSET to the resonator,
leading to an effective negative damping. For a superconducting
stripline resonator, the quality factor can be high enough that the
negative damping dominates and the resonator is driven into an
oscillating state even in the absence of an external drive
\cite{ria,bc,Hauss}. This has recently been observed for a
superconducting stripline resonator coupled to an SSET, and the
field transmitted through the resonator measured \cite{Tsai07}. When
the resonator is in this oscillating state, the SSET has a periodic
but not harmonic response, with the details of its dynamics
depending rather sensitively on the amplitude of the resonator
oscillation. In order to probe this behavior, we must go beyond the
time-averaged mean field behaviour \cite{riha} and  investigate the
correlations present in the system.

With a stripline resonator, as opposed to a mechanical one, there
are two independent probes of the system - the field in the
superconducting stripline, and the current that passes through the
SSET. This then allows the investigation of the system by two
different experimental methods. The noise properties of the
SSET-cavity system are transferred to both the current and reflected
field, so we can consider these to be probing the noise present in
our system. The effect of the thermal noise in both the SSET and the
cavity can be made very small, so the noise detected largely arises
from the finite level structure of the SSET, and is in that sense is
a purely quantum noise.

In this paper, we consider an SSET that is coupled to a resonator
that is driven into an oscillating state. We calculate the charge
noise on the SSET and the effect this has on the cavity field. We
also consider the case when the resonator's motion depends on the
effective damping and noise arising from the SSET, using a linear
response approach \cite{Clerk}. We go on to show how the response of
the system at the sidebands could be used to detect small changes in
the driving amplitude. Although we focus our analysis on the case of
a superconducting stripline, which has a higher oscillation
frequency, our main results are also valid for the lower frequency
mechanical resonators. In our calculation, we take advantage of the
fact that the resonator oscillations change on a timescale that is
slow compared to the SSET timescales, which is valid as long as the
resonator damping and coupling to the SSET are weak. This means that
the SSET response is essentially the response to a purely harmonic
drive. This approach has previously been used for both mechanical
and optomechanical systems \cite{riha,MHG}. However, by including a
fluctuating Langevin term, we can go beyond the previous mean field
results to calculate the noise. The effect of this response on the
resonator can then be calculated.

In Sec. \ref{sec:equations} we introduce our model of the system and
Langevin equations for the resonator field and the SSET variables.
In Sec. \ref{sec:solveL02} we introduce the main approximations made
and solve the Langevin equations for the SSET under the influence of
the periodic driving provided by the resonator. Section
\ref{sec:noisech} describes the frequency spectrum of the charge on
the SSET and gives approximate expressions valid when the resonator
frequency is large. In Sec. \ref{sec:cavity} we show how the charge
dynamics affect the field in the cavity. We derive expressions for
the effective SSET damping and temperature of the resonator, going
beyond previously published results to derive expressions valid for
all resonator frequencies. Section \ref{sec:detect} describes how
the response of the system at multiples of the resonator frequency
could be used to detect small changes in the driving amplitude and
Section \ref{sec:output} describes how the field in the cavity could
be measured through a transmission line coupled to the cavity. In
Sec. \ref{sec:conclusions} we present our conclusions and in
appendix \ref{sec:apLV} we present more details of the derivation of
the Langevin equations for the SSET.

\section{Master Equation and Langevin Equations}
\label{sec:equations} The master equation for a superconducting
single electron transistor biased at the Josephson quasiparticle
resonance and capacitively coupled to a resonator \cite{bia,ria},
\begin{equation}
\dot{\rho}={\mathcal L}\rho=-\frac{i}{\hbar}[H_{\rm
co},\rho]+\mathcal{L}_{leads}\rho+\mathcal{L}_{damping}\rho.
\label{master}
\end{equation}
consists of a coherent part, described by the Hamiltonian, $H_{\rm
co}$, together with two dissipative terms $\mathcal{L}_{leads}$ and
$\mathcal{L}_{damping}$ which describe quasiparticle decay from the
island and the surroundings of the resonator respectively. The
effective Hamiltonian is given by,
\begin{eqnarray}
H_{\rm co}&=&-\hbar\Delta
\sigma_{22}-\hbar\epsilon_J\left(\sigma_{02}+\sigma_{20}\right)\nonumber
\\&&+\hbar\omega_0a^{\dag}a+ \hbar A_D(ae^{i\omega_Dt}+a^\dag
e^{-i\omega_Dt})\nonumber \\&&-
\hbar\omega_0\frac{x_s}{2x_q}(a+a^\dag)\left(\sigma_{11}+2\sigma_{22}\right),
\label{Ham}
\end{eqnarray}
where the operators $\sigma_{ij}=|i\rangle\langle j|$ represent
operators on the SSET island charge states $|i\rangle$, $\Delta$ is
the energy difference between states $|2\rangle$ and $|0\rangle$ and
$\epsilon_J$ is the Josephson energy of the superconductor.  The
frequency of the resonator is $\omega_0$ and $A_D,\omega_D$ give the
strength and frequency of the external driving respectively. The
resonator-SSET coupling is described by the parameter $x_s$ which
measures the shift in the equilibrium field of the resonator brought
about by adding a single electronic charge to the SSET
island\,\cite{bia}, and
$x_q=(\hbar/2\epsilon_0\omega_0)^\frac{1}{2}$. The tunneling of
quasiparticles from the island is described by
\begin{eqnarray}
\mathcal{L}_{leads}\rho&=&-\frac{\Gamma}{2}\left[\left\{\sigma_{22}
+\sigma_{11},\rho\right\}_+\right.\\&&\left.-2\left(\sigma_{12}+\sigma_{01}\right)\rho\left(\sigma_{21}+\sigma_{10}\right)\right],
\nonumber
\end{eqnarray}
where we have neglected the position dependence of the tunnel rates
as being of lesser importance than the coherent coupling
\cite{ria,riha}. This simplification means that the dissipation
takes a Lindblad form, and also means that the master equation is
essentially equivalent to that of a resonator coupled to a double
quantum dot. The dissipation and fluctuations arising from the
resonator's surroundings are described by the usual quantum optical
expression \cite{WM},
\begin{eqnarray}
\mathcal{L}_{damping}\rho&=&-\frac{\gamma_{ex}}{2}(\overline{n}+1)\left(
a^{\dagger}a\rho+\rho a^{\dagger}a-2a\rho a^{\dagger}\right)
 \\ &&-\frac{\gamma_{ex}}{2}\overline{n}\left(
aa^{\dagger}\rho+\rho aa^{\dagger}-2a^{\dagger}\rho
a\right)\nonumber\label{eq:qodiss}
\end{eqnarray}
where this form for the dissipation is valid as long as  the
dynamics of the system are slow compared to the correlation time of
the bath.

An alternative description of the system is given by a set of
Langevin equations describing the coupled system. This gives an
equivalent description of the first and second moments of the
system, which is what is required for a noise calculation. We shall
see later that this Langevin form is convenient for dealing with the
external drive, as a transformation can be made which allows us to
solve the equation analytically.

The Langevin equations are equal to the semiclassical equations plus
a fluctuating noise term. For the resonator operator $a$ we have,
\begin{eqnarray}
\dot{a}&=& -i\omega_0a -\frac{\gamma_{ex}}{2}a -iA_De^{-i\omega_Dt}+\eta_a\nonumber\\
&& +i\omega_0\frac{x_s}{2x_q}(\sigma_{11}+2\sigma_{22})
\label{eq:fieldLGV}
\end{eqnarray}
where $\eta_a$ represents the standard white noise term on the
resonator defined by $\av{\eta_a(t)}=0$ , $\av{\eta_a^\dag(t)
\eta_a(t')}=\delta(t-t')\gamma_{ex}\bar{n}$ and $\av{\eta_a(t)
\eta_a^\dag(t')}=\delta(t-t')\gamma_{ex}(\bar{n}+1)$. This couples
to the equations for the SSET charge operators
$\sigma_{ij}=\ket{i}\bra{j}$ \cite{riha}. The Langevin equations for
the projection operators are given by,
\begin{eqnarray}
\dot{\sigma}_{00}(t) &=& i \epsilon_J(\sigma_{02}(t)-\sigma_{20}(t))
+\Gamma \sigma_{11}(t)+\eta_{00}(t). \label{eq:Lcs1}\\
\dot{\sigma}_{11}(t)
&=&\Gamma \sigma_{22}(t)-\Gamma \sigma_{11}(t)+\eta_{11}(t)\label{eq:Lcs2}\\
\dot{\sigma}_{22}(t)&=&  -i \epsilon_J
(\sigma_{02}(t)-\sigma_{20}(t)) -\Gamma \sigma_{22}(t)+\eta_{22}(t)
\label{eq:Lcs3}
\end{eqnarray}
which also couple to the equation of  motion for the off-diagonal
terms describing the coherence between levels $\ket{0}$ and
$\ket{2}$,
\begin{eqnarray}
\dot{\sigma}_{02}(t)&=&-i\epsilon_J(\sigma_{22}(t)-\sigma_{00}(t))\!-\!\frac{\Gamma}{2}\sigma_{02}(t)\label{eq:L02}\\
&& +i(\Delta\! +\! \frac{\omega_0x_s}{x_q}
(a(t)\!+a^\dag(t)))\sigma_{02}(t) +\eta_{02}(t).\nonumber
\end{eqnarray}
 We need
to find the properties of the correlators for the SSET noise
operators $\eta_{ij}$. The Langevin equations, including the
properties of the noise operators, can be easily derived from the
master equation by
requiring that the two forms give the same equations of motion for
the second moments of the operators, determining the correlation
properties of $\eta_{ij}$ (see appendix \ref{sec:apLV}). For
example, for the operator representing the noise on $\sigma_{02}$ we
have,
\begin{eqnarray}
\av{\eta_{02}(t)\eta_{20}(t')}&=&\Gamma(\av{\sigma_{00}(t)}+\av{\sigma_{11}(t)})\delta(t-t').
\label{eq:n02n20}
\end{eqnarray}
There are a few points to note about the noise operators for the
SSET. Firstly, in the limit $k_BT\ll \hbar \epsilon_J$, the SSET
experiences no direct thermal noise. The ``noise" terms
$\eta_{02}(t)$ therefore arise solely from the finite level
structure of the SSET and the resulting commutation relations, and
in that sense can be considered to be purely quantum noise.
Secondly, we are giving an approximate treatment of this quantum
noise by requiring that the first and second order moments
calculated by the quantum Langevin equations are equivalent to the
first and second order moments as determined by the master equation.
In general these would not suffice to determine all the moments in
the system. Finally, we note is that we are describing a system
under periodic driving which will therefore tend to a periodic
behavior in the long time limit, rather than a fixed point. In a
finite level system, the correlators involve the average values of
the operators such as $\av{\sigma_{11}(t)}$ which depend on $t$ in a
driven system. This means that we have time-dependent noise
correlators for the SSET noise operators.

The equations for the operators $\sigma_{12}$,$\sigma_{10}$ decouple
from the others, so Eqs. (\ref{eq:Lcs1}-\ref{eq:L02}), along with
their correlators, completely determine the dynamics of the
SSET-resonator system, up to second order.

\section{Solving the $\sigma_{02}$ Langevin Equation}
\label{sec:solveL02}

In this section we solve the Langevin equation (Eq. (\ref{eq:L02}))
for the off-diagonal charge operator $\sigma_{02}$. In order to
progress we make two assumptions: first that the Josephson energy is
rather small, and second that the amplitude of the resonator changes
slowly compared to the incoherent dynamics of $\sigma_{02}$, {i.e.}
that the total damping due to both the resonator environment and the
SSET $\gamma_{T}=\gamma_{ex}+\gamma_{SS}$ is much smaller than the
quasiparticle decay rate $\Gamma$.

 In the limit that the Josephson
energy is much weaker than the quasiparticle decay, $\epsilon_J\ll
 \Gamma$, the occupation of the charge states $\sigma_{11},\sigma_{22} \ll
 1$, and the equation of motion for $\sigma_{02}$ decouples from the
other charge equations.
We are assuming the resonator amplitude can be treated as constant
on timescales relevant to the SSET dynamics, an approximation which
has proved useful in related mechanical and optomechanical systems
\cite{riha,MHG}. Our derivation closely follows these methods, but
also incorporates the fluctuations described by the $\eta$ terms. We
replace the term describing the field in Eq. (\ref{eq:L02}) with a
cosine oscillation at the driving frequency with magnitude $A$
{i.e.} $(a(t)+a^\dag(t))\approx A\cos (\omega_D t) $. For the case
of weak back-action damping, the amplitude is simply given by
$A=A_D/((\omega_0-\omega_D)^2+\gamma_{ex}^2)^{1/2}$. When the back
action damping is significant, the amplitude must be determined
self-consistently - an oscillation of amplitude $A$ will be stable
if the resulting total damping is zero.

This then means that the field simply appears as an harmonic drive
acting on the SSET, and Eq. (\ref{eq:L02}) becomes,
\begin{eqnarray}
\dot{\sigma}_{02}(t)\!\!\!&=&\!i\epsilon_J\!-\!\left(\frac{\Gamma}{2}
\!-\!i\Delta\! -\!
i \frac{\omega_0x_s}{x_q} A \cos \omega_D t\right)\sigma_{02}(t)\!+\eta_{02}(t),\nonumber\\
\label{L02s}
\end{eqnarray}
where the first term on the right hand side multiplies an implicit
identity operator.

If we make a transformation to eliminate the driving term, $\tilde
\sigma_{02}=\sigma_{02}e^{-iz \sin \omega_D t }$, we can then find
the Fourier transform of the transformed operator
$\tilde\sigma_{02}$ in terms of Bessel functions of the first kind,
$J_n(z)$, where $z=(\omega_0x_s A)/(\omega_D x_q)$,
%
\begin{eqnarray}
\tilde{\sigma}_{02}(\omega_F)&=& \frac{i\epsilon_J\sum\limits_n
\delta(\omega_F-\omega_Dn)J_n\left(-z\right)}{\Gamma/2+i(\omega_F-\Delta)
}\nonumber\\&& +\frac{\frac{1}{2\pi}\int\limits_{-\infty}^\infty
e^{-i\omega_F t}e^{-iz \sin \omega_D t }\eta_{02}(t) \rd
t}{\Gamma/2+i(\omega_F-\Delta) } \label{eq:FT02trans}
\end{eqnarray}
and we see that $\tilde \sigma_{02}$ consists of a systematic
response to the drive at multiples of the driving frequency
$\omega_D$, plus a noise term.

 Equation (\ref{eq:FT02trans}) gives the
Fourier component in the transformed picture, so we
%
convert back to the untransformed picture to obtain,
\begin{eqnarray}
{\sigma}_{02}(\omega_F)&=&  \sum\limits_{n,n'}{\frac{i\epsilon_J
\delta(\omega_F-\omega_Dn')J_{n'-n}\left({\scriptstyle-}z\right)
 J_{n}\left ( z \right)}{\Gamma/2+i(\omega_D(n'-n) -\Delta)
 }}\nonumber\\
 &+&\sum\limits_{n}\frac{\int\limits_{-\infty}^\infty
 e^{-i(\omega_Ft-\omega_Dnt+z \sin \omega_D t) }\eta_{02}(t) J_n\left ( z \right)\rd
t}{2\pi(\Gamma/2+i(\omega_F -\omega_D n-\Delta)) },\nonumber\\
\label{eq:FTp02}
\end{eqnarray}
which gives an expression for the Fourier transform of
$\sigma_{02}(t)$ consisting of a mean field and noise-induced term.
From this expression we can go on to calculate the spectrum of the
charge on the SSET and hence of the cavity field.

\section{Noise on the SSET} \label{sec:noisech}
 The frequency
spectrum of two fluctuating terms $f(t),g(t)$ is defined by,
\begin{eqnarray}
S_{fg}(\omega,t)&=& \int\limits_{-\infty}^{\infty} \rd \tau e^{i
\omega \tau} \av{f(t+\tau)g(t)}
\nonumber\\
& =&2\pi\int\limits_{-\infty}^{\infty} \rd \omega_F e^{i
(\omega_F+\omega) t} \av{f(\omega)g(\omega_F)} \label{eq:FTdef1}
\end{eqnarray}
where in the second line we have written the spectrum in terms of
the Fourier transforms of the individual functions. Note that as
defined, the spectrum includes correlations due to the systematic
motion of the terms as well as due to the fluctuations. If the
functions $f,g$ are periodic then the above expression retains a
periodic time dependence. We therefore define a spectrum averaged
over a single driving period,
$S_{fg}(\omega)=\frac{\omega_D}{2\pi}\int\limits_{-\pi/\omega_D}^{\pi/\omega_D}
\rd t S_{fg}(\omega,t)$, which is what we would expect to observe in
experiment. Inserting Eq. (\ref{eq:FTp02}) into this expression and
noting that in the small-$\epsilon_J$ limit we can approximate the
time dependent correlator given in Eq. (\ref{eq:n02n20}) as
$\av{\eta_{02}(t)\eta_{20}(t')}\approx\Gamma\delta(t-t')$, gives,
\begin{eqnarray} \label{eq:S02s}
S_{\sigma_{02}\sigma_{20}}(\omega)\!&=&\!
\sum\limits_{n}2\pi\delta(\omega\!-\!\omega_Dn)\left|\sum\limits_{n'}
\frac{\epsilon_J J_{n'}\left(\text{-}z\right)
 J_{n\text{-}n'}\left( z \right) }{\frac{\Gamma}{2}+i(\omega_Dn'-\Delta)
 }\right|^2\nonumber\\
 &&+\sum\limits_{n}\frac{\Gamma J_n\left(
z \right)^2}{(\omega -\omega_D n-\Delta)^2 +\Gamma^2/4}.
\end{eqnarray}
The first term corresponds to the correlations arising from the
systematic response of the resonator to the driving force. For a
purely monochromatic drive, the peaks are $\delta-$functions, but in
reality the $\delta-$functions would be replaced by peaks with total
power unity and a width determined by the linewidth of the driving.
The second term describes the additional noise due to the finite
level structure.

We have calculated the noise on the off-diagonal element of the SSET
density matrix, $\sigma_{02}$. However, this does not enter directly
into the current or cavity noise. Instead, we have to calculate the
frequency spectrum of the \emph{total charge} on the SSET,
$\sigma_{cc}=\sigma_{11}+2\sigma_{22}$.

\subsection{Systematic Charge Oscillations} \label{sec:syschg}

Fourier transforming Eqs. (\ref{eq:Lcs1}-\ref{eq:Lcs3}) gives an
expression for the charge in terms of $\sigma_{02}(\omega_F)$.
Inserting Eq. (\ref{eq:S02s}) into this gives an expression for the
charge consisting of a systematic part, and a fluctuating part
containing the noise operators $\eta_{02},\eta_{11}$ and
$\eta_{22}$. We now go on to calculate the fluctuation spectrum of
the charge due to the systematic and noisy terms.

The systematic component of the charge, $\sigma_{02}^S(\omega_F)$ is
given by,
\begin{eqnarray}
\sigma^S_{cc}(\omega_F)&=&\frac{(3\Gamma
+2i\omega_F)\epsilon_J^2}{(\Gamma+i\omega_F)^2}\sum\limits_{n,n'}J_{n'}\left(z\right)\delta(\omega_F-\omega_Dn)\nonumber\\
&&\times\left(
\frac{J_{n'+n}\left(z\right)}{\frac{\Gamma}{2}-i(\omega_Dn'\!+\!\Delta)}
+\frac{J_{n'\text{-}n}\left(z
\right)}{\frac{\Gamma}{2}+i(\omega_Dn'\!+\!\Delta)}
\right),\nonumber\\\label{eq:p11p22FTS}
\end{eqnarray}
and we see that the mean field response of the charge is a series of
$\delta-$peaks at multiples of $\omega_D$.

The systematic motion of the charge will show up in the time
correlation function and hence the fluctuation spectrum. We insert
Eq. (\ref{eq:p11p22FTS}) into Eq. (\ref{eq:FTdef1}) to obtain the
part of the charge spectrum due to systematic evolution,
\begin{eqnarray}
S^S_{cc}(\omega) &=& \sum\limits_n 2\pi \delta(\omega+n\omega_D)
\epsilon_J^4
\frac{9\Gamma^2+4\omega_D^2n^2}{(\Gamma^2+\omega_D^2n^2)^2}
\\
&\times& \!\left| \sum_{n'} \left(
\frac{J_{n'}\!\left(z\right)J_{n+n'}\left(z\right)}{\frac{\Gamma}{2}-i(\omega_Dn'\!+\!\Delta)
}
+\frac{J_{n'}\!\left(z\right)J_{n'\text{-}n}\left(z\right)}{\frac{\Gamma}{2}+i(\omega_Dn'\!+\!\Delta)
 } \right)\right|^2\nonumber
\end{eqnarray}
For $\Gamma,\Delta \ll \omega_D$, this expression is dominated by
the $n'=0$ term in the sum, and the systematic noise reduces to,
\begin{eqnarray}
S^S_{cc}(\omega) &\approx& \sum\limits_n
2\pi\delta(\omega+n\omega_D) \epsilon_J^4 \frac{4}{(\omega_Dn)^2}
J_{0}\left(z\right)^2\nonumber\\
&&\times \left| \left(
\frac{J_{n}\left(z\right)}{\frac{\Gamma}{2}-i\Delta }
+\frac{J_{-n}\left(z\right)}{\frac{\Gamma}{2}+i\Delta
 } \right)\right|^2
\end{eqnarray}

The symmetry properties of the Bessel functions then mean that the
bessel functions will either add or cancel depending on whether $n$
is odd or even, so separating out the two cases we find,
\begin{eqnarray}
S^{S}_{cc}(\omega) &\approx& \frac{ 4\epsilon_J^4
J_{0}\left(z\right)^2}{(\Delta^2 +\frac{\Gamma^2}{4})^2}
\sum\limits_{n} \frac{2\pi \delta(\omega+n\omega_D) }{(\omega_Dn)^2}
J_{n}\left(z\right)^2 \nonumber\\
&& \times\left(\Gamma^2 \frac{(1+(-1)^n)}{2} +4\Delta^2
\frac{(1-(-1)^n)}{2}\right)  \nonumber\\ \label{eq:sysnoise}
\end{eqnarray}

\subsection{Charge Spectrum Due to Fluctuations}

\begin{figure}\centering{
\includegraphics[width=8cm]{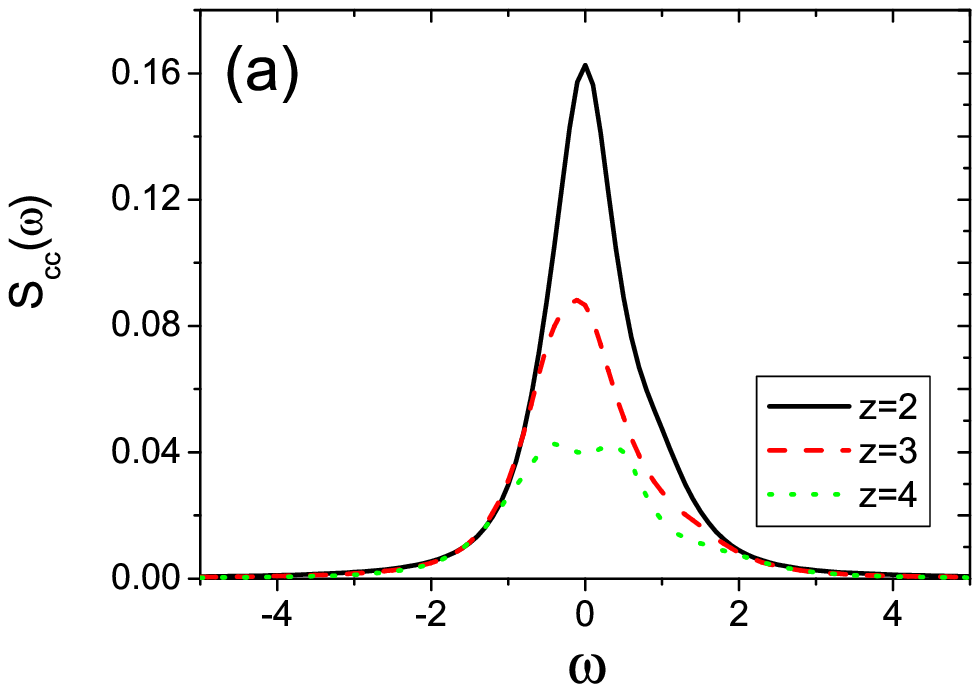}\\
\vspace{-0.75cm}
\includegraphics[width=8cm]{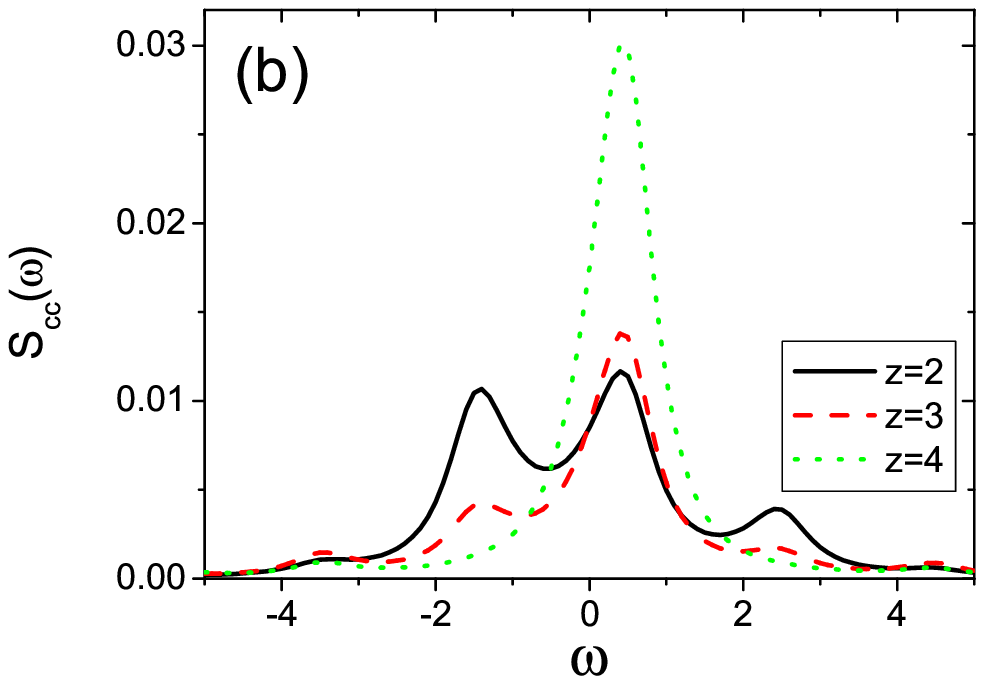}\\
\vspace{-0.75cm}
\includegraphics[width=8cm]{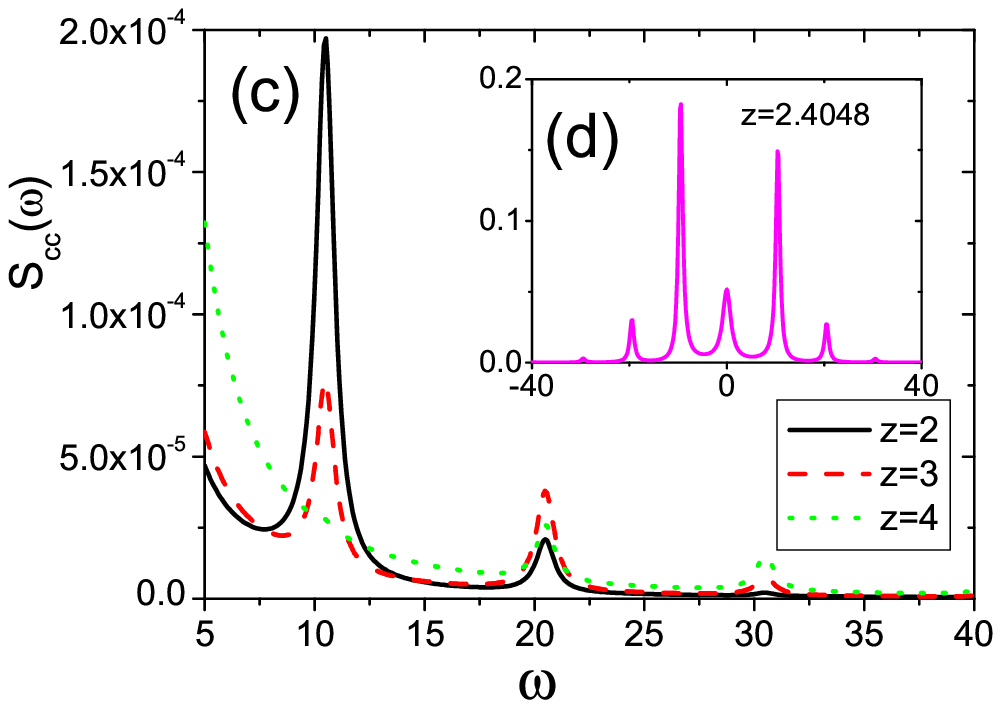}}
\caption{(color online). Charge noise as a function of $\omega$ for
different driving frequencies $\omega_D=(a)0.5, (b)2, (c)10$
relative to the quasiparticle decay $\Gamma$. The different curves
show different driving amplitudes  $z=2,3,4$. The inset (d) in the
$\omega_D=10$ plot shows that the large central peak is highly
suppressed at $J_0(z)=0$, here at $z=2.4048$. The other parameters
are $\epsilon_J=1/16, \Gamma=1, \Delta=-0.5,$} \label{fig:noisew}
\end{figure}

As well as the systematic response to the driving force, the charge
spectrum also contains a part due to fluctuations. The Fourier
transform of the charge $\sigma_{cc}=\sigma_{11}+2\sigma_{22}$
contains the fluctuating terms corresponding to both the diagonal
and off-diagonal operators, $\eta_{11},\eta_{22},\eta_{02}$.
Inserting these into Eq. (\ref{eq:FTdef1}), we write the
fluctuation-induced part of the charge spectrum as,
\begin{eqnarray}
S^\eta_{cc}&=&S^{diag}_{cc}+S^{\eta_{02},\eta_{20}}_{cc}+S^{cross}_{cc}
\end{eqnarray}
where $S^{diag}_{cc}$ corresponds to the part arising from
fluctuations on the diagonal operators $\eta_{11},\eta_{22}$, the
term $S^{02}_{cc}$ arises from $\eta_{02}$, and $S^{cross}_{cc}$
from the correlations between these. Note that while
$\av{\eta_{02}(t)\eta_{20}(t')}$ can be approximated to a constant
in the small-$\epsilon_J$ limit, other correlators give, for
example,
$\av{\eta_{22}(t)\eta_{22}(t')}=\Gamma\av{\sigma_{22}(t)}\delta(t-t')$,
{i.e.} we have a time dependent function multiplied by a delta
function. As $\av{\sigma_{22}(t)}$ is simply the systematic part of
$\sigma_{22}$, we can insert the Fourier transform of this into our
calculation. After some algebra, we find,
\begin{eqnarray}
S_{cc}^{diag}(\omega)&=&
\frac{5\Gamma^2+2\omega^2}{(\Gamma^2+\omega^2)^2}\sum\limits_n\frac{J_{n}\left(z\right)^2\Gamma\epsilon_J^2}{\frac{\Gamma^2}{4}+(\omega_D
n + \Delta)^2}
\label{eq:noisesum2}\\
S_{cc}^{\eta_{02},\eta_{20}}(\omega)
&=&\frac{9\Gamma^2+4\omega^2}{(\Gamma^2+\omega^2)^2}\sum\limits_n
\frac{J_{n}\left(z \right)^2
\Gamma\epsilon_J^2}{\frac{\Gamma^2}{4}+(\omega-\omega_D n -
\Delta)^2} \nonumber \\\label{eq:noisesum1}.
\end{eqnarray}
The expression for the cross terms is rather unwieldy, but can be
written as,
\begin{eqnarray}
S_{c,c}^{cr}(\omega)&=&\Re\Bigg\{\frac{12\Gamma^2+4\omega^2+2i\omega\Gamma}{(\Gamma^2+\omega^2)^2}
\\
&\times&\sum\limits_n \frac{J_{n}\left(z \right)^2
\Gamma\epsilon_J^2}{(\frac{\Gamma}{2}-i(\omega_Dn\!-\!\Delta))
(\frac{\Gamma}{2}+i(\omega-\!\omega_Dn-\!\Delta))}\Bigg\}\nonumber
\label{eq:noisesum3}
\end{eqnarray}
 In Fig. \ref{fig:noisew} we plot the
charge noise as a function of $\omega$, and see it consists of a
series of peaks (of width $\sim\Gamma$) at integer multiples of the
driving frequency on top of a background peak centered on
$\omega=0$.
It is these sideband peaks that give information about the driving
frequency and amplitude, and so we would like to find some simple
expressions that tell us if these peaks can be observed. The peaks
are more pronounced in the limit that the resonator frequency is
larger than the quasiparticle decay rate, and this is what we would
expect to be the case experimentally for superconducting stripline
resonators.

 The spectrum simplifies
considerably in the large $\omega_D$ limit. The diagonal term, Eq.
(\ref{eq:noisesum2}), consists of a peak at $\omega=0$, multiplied
by a sum that is independent of $\omega$. We can approximate the sum
by the $n=0$ term in the $\Gamma,\Delta\ll\omega_D$ limit to obtain,

The other terms describe a similar peak at $\omega=D$, multiplied by
a series of peaks. We wish to find an expression for the heights of
these peaks. In the $\Gamma,\Delta\ll\omega_D$ limit, we find that
the heights of the noise terms at integer values of the driving
frequency are,
\begin{eqnarray}
 S^{\eta_{02},\eta_{20}}_{cc}(\omega_Dn)
 &\approx&\frac{4 J_n(z)^2}{(\omega_Dn)^2}\frac{\Gamma\epsilon_J^2}{\frac{\Gamma^2}{4}+\Delta^2}\nonumber\\
  S^{cr}_{cc}(\omega_Dn)&\approx&-\frac{4 J_n(z)^2\Delta}{(\omega_Dn)^3}\frac{\Gamma\epsilon_J^2}{\frac{\Gamma^2}{4}+\Delta^2}
  \end{eqnarray}
and we note that the cross terms are negligible in this limit and
can be neglected. Thus we can write the height of the charge peaks
as
\begin{eqnarray}
 S^{\eta}_{cc}(\omega_Dn)
 &\approx&\frac{(4
 J_n(z)^2+2J_0(z)^2)}{2\pi(\omega_Dn)^2}\frac{\Gamma\epsilon_J^2}{\frac{\Gamma^2}{4}+\Delta^2} . \label{eq:approxonpk}
\end{eqnarray}
The visibility of the peaks depends on the contrast between the
noise at the peaks and the noise between the peaks
\cite{Doiron2QPC}.

\begin{figure}\centering{
\includegraphics[width=8cm]{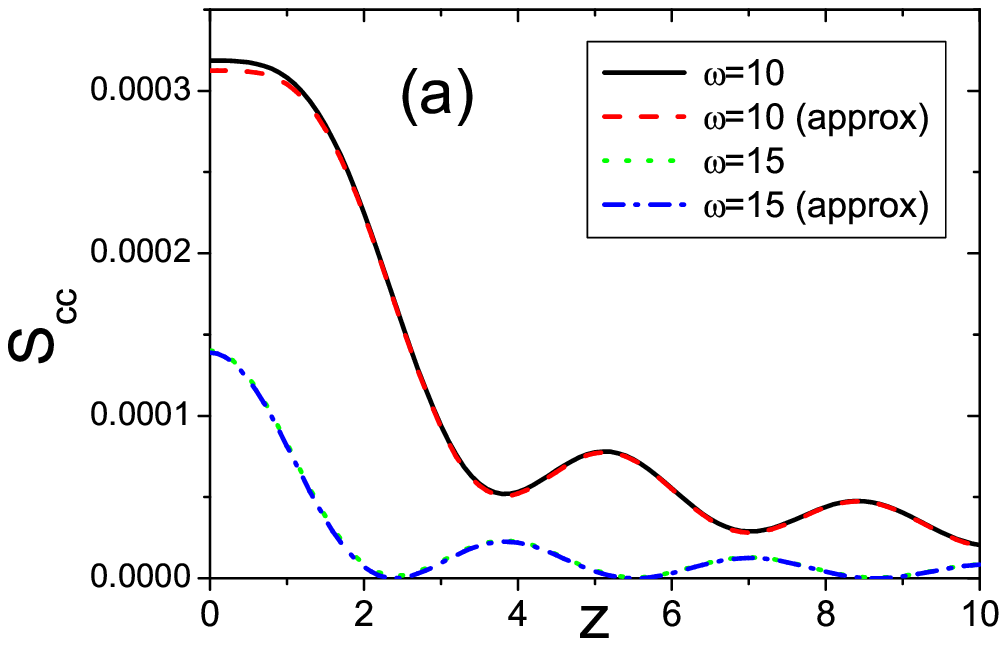}\\
\vspace{-0.75cm}
\includegraphics[width=8cm]{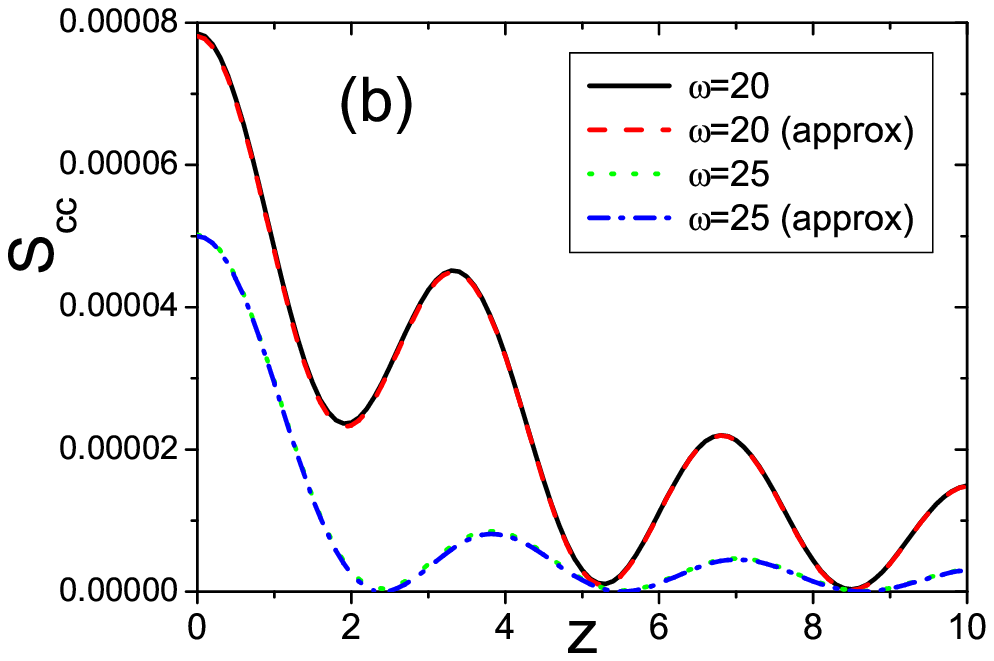}\\
\vspace{-0.75cm}
\includegraphics[width=8cm]{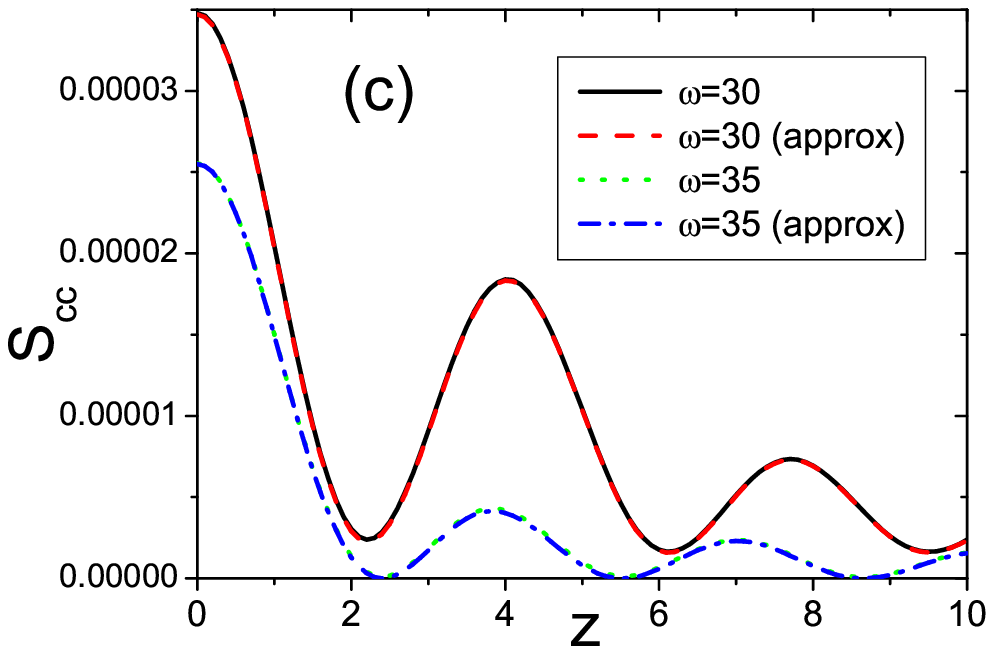}}
\caption{(color online). Comparison of the heights of the peaks at
$\omega=\omega_0n$ and the troughs at $\omega=\omega_0(n+1/2)$, for
$n=(a)1, (b)2, (c)3$ respectively. Both the full calculated value
and the $\Gamma,\Delta \ll \omega_0$ approximation are plotted.
Parameters: $\omega_0=10, \Gamma=1, \Delta=0,\epsilon_J=1/16,$}
\label{fig:onoffpk}
\end{figure}

Between the peaks, two terms in the sum are relevant which gives, to
leading order,
\begin{eqnarray}
 S^{\eta}_{cc}(\omega_D(n+\frac{1}{2}))&\approx&\frac{2J_0(z)^2}{\omega_D^2(n+\frac{1}{2})^2}
 \frac{\Gamma\epsilon_J^2}{\frac{\Gamma^2}{4}+\Delta^2} \label{eq:approxoffpk}
\end{eqnarray}
The ratio reaches a minimum when $J_n(z)=0$, at which point we see
the ratio is $(n+\frac{1}{2})^2/n^2$. The ratio becomes large when
$J_0(z)=0$, and is of order
$\omega_D^2n^2/(\frac{\Gamma^2}{4}+\Delta^2)$. Examples of the on-
and off-peak heights as a function of the driving are shown in Fig.
(\ref{fig:onoffpk}), along with the approximations to these given in
Eqs. (\ref{eq:approxonpk}-\ref{eq:approxoffpk}).

\begin{eqnarray}
 S^{diags}_{cc}(\omega)&\approx&\frac{5\Gamma^2+2\omega^2}{(\Gamma^2+\omega^2)^2}\frac{J_0(z)^2\Gamma\epsilon_J^2}{\frac{\Gamma^2}{4}+\Delta^2}\nonumber\\
  \end{eqnarray}

\section{Cavity Field} \label{sec:cavity}

We now consider the effect of the SSET on the cavity. We recall the
Langevin equation for the cavity field,
\begin{eqnarray}
\dot{a}&=& -i\omega_0a -\frac{\gamma_{ex}}{2}a -iA_De^{-i\omega_Dt}+\eta_a\nonumber\\
&& +i\omega_0\frac{x_s}{2x_q}(\sigma_{11}+2\sigma_{22})
\end{eqnarray}
where $A_D$ represents the amplitude of the classical driving field.

If we neglect the back-action as weak, then the steady state
amplitude $A=|a|$ of the resonator is simply given by,
\begin{eqnarray}
A&=&
\frac{A_D}{\sqrt{(\omega_0-\omega_D)^2+\frac{\gamma_{ex}^2}{4}}}.
\end{eqnarray}
We can now use this amplitude to calculate the behavior of the SSET.
If the back-action of the SSET on the resonator is weak, then the
oscillations of the resonator at the driving frequency will be
relatively unaffected and hence the drive the SSET field feels will
be unchanged. However, the SSET does not simply respond at the
driving frequency, but has a systematic response at multiples of
$\omega_D$, and will also act as an additional source of noise.

Assuming that the response of the SSET at the driving frequency is
unaffected by the back-action, the Fourier transform of $a$ is given
by,
\begin{eqnarray}
a(\omega_F)&=& \frac{-iA_D\delta(\omega_F+\omega_D)+i\omega_0\frac{x_s}{2x_q}\sigma_{cc}^S(\omega_F)}{i(\omega_F+\omega_0)+\frac{\gamma_{ex}}{2}}\nonumber\\
&& +\frac{\frac{1}{2\pi}\int e^{-i(\omega_D+\omega_F)t}\eta_a(t)\rd
t
+i\omega_0\frac{x_s}{2x_q}\sigma_{cc}^\eta(\omega_F)}{i(\omega_F+\omega_0)+\frac{\gamma_{ex}}{2}}\nonumber\\\label{eq:resnoisenba}
\end{eqnarray}
where the first line describes the systematic motion of the cavity
and the second gives the noise. We now see that our assumption that
the change in the oscillation amplitude ({i.e.} the change of
$a(\omega_D)$) is negligible will be justified whenever
$A_D\gg\omega_0\frac{x_s}{x_q}\sigma_{cc}^S(-\omega_D)$. At other
frequencies the SSET may have a significant effect on the resonator.
The systematic response of the resonator at frequency $\omega$ is
just the charge response at that frequency multiplied by a prefactor
of magnitude $\omega_0/((\omega-\omega_0)^2+\gamma_{ex}^2)^2$.

We can now write the cavity noise as a function of the charge noise,
\begin{eqnarray}
S_{a^\dag,a}(\omega)=\frac{\bar{n}\gamma_{ex}+\frac{\omega_0^2x_s^2}{4xq^2}(S^S_{c,c}(\omega)+S^\eta_{c,c}(\omega))}
{(\omega-\omega_{0})^2+\frac{\gamma_{ex}^2}{4}} \label{eq:resnoise}
\end{eqnarray}
with the systematic and noise-induced parts of the charge spectrum
calculated driving amplitude $A$. We see that the charge noise
spectrum appears directly in the expression for the cavity noise,
and so the sidebands discussed in Sec. \ref{sec:noisech} should also
be present. Although the size of the peaks is suppressed, we find
that (for $\bar{n}=0$), the ratio of the on- and off-peak noise is
unchanged.

\subsection{Back-action damping and temperature} \label{sec:damp}

In the preceding section, we assumed that the back-action was weak
enough that the SSET damping could be neglected. However, it is well
known that the effect of the back action on the dynamics of the
resonator can be significant. In particular, as well as providing an
additional source of noise for the resonator, the SSET can also act
to provide an additional source of damping
\cite{bia,ria,riha,bc,SSET1}. In these situations, this effective
damping will have a significant effect on the noise properties of
the resonator even when the resonator is strongly driven.
Furthermore, the SSET can cause the total resonator damping of the
resonator to become negative, and hence the resonator can be driven
into a self-oscillating laser-like state even in the absence of
external driving \cite{bc,SSET1,ria,riha}. In this section we review
how the systematic response of the SSET can act as an
amplitude-dependent damping of the resonator, and present some
simple analytic approximations before describing how this influences
the cavity noise spectrum. We show that the calculation of charge
noise presented in Sec. \ref{sec:noisech} can be used to generalize
the previously known expressions for the effective damping and
temperature of a slow ($\omega_0\ll\Gamma$) resonator to arbitrary
frequency using a linear-response-like approach \cite{Clerk}.

\begin{figure}\centering{
\includegraphics[width=8cm]{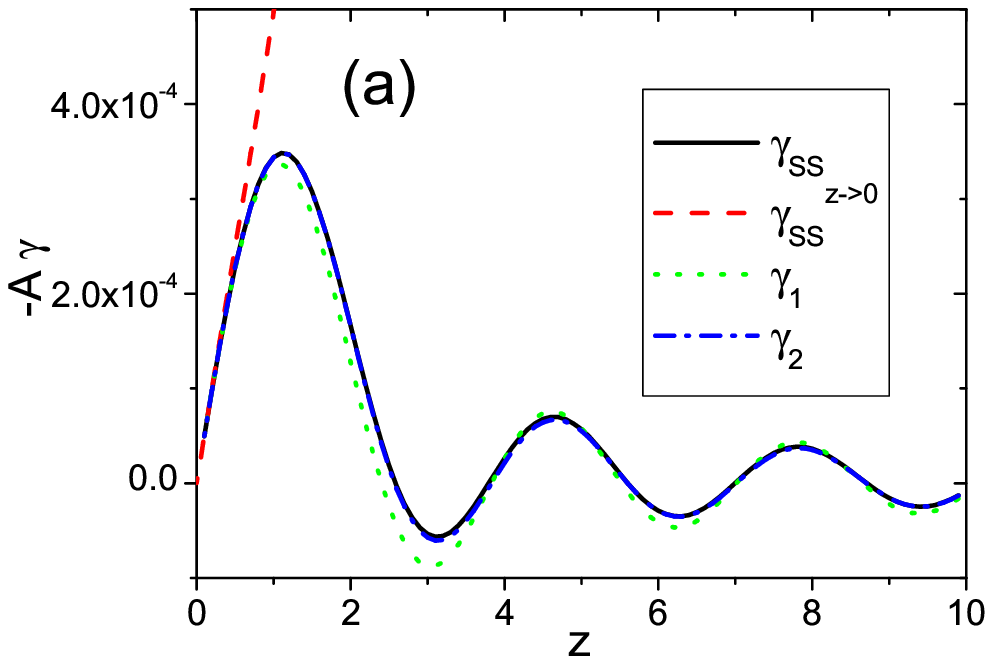}\\
\vspace{-0.75cm}
\includegraphics[width=8cm]{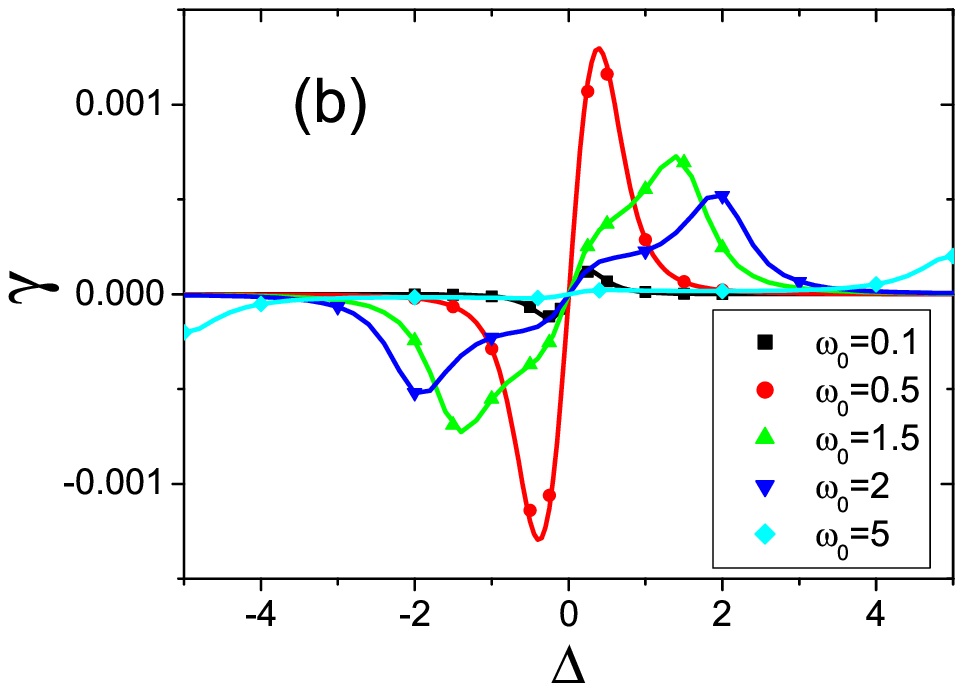}}
\caption{(color online). Plot of the SSET damping. Plot (a) shows
non-linear damping, with $-\gamma_{SS}A$ plotted against $A$.
Plotted are the full expression for the damping, $\gamma_{SS}$, the
linear damping, $\gamma_{lin}$, and approximations for the nonlinear
damping using one ($\gamma_1$) or two ($\gamma_2$) terms from Eq.
(\ref{eq:nldampap}), with $\kappa=0.1, \Delta=-0.2, \epsilon_J=0.05,
\Gamma=1$. Plot (b) shows the linear damping over a range of
resonator frequencies, where the points are the damping calculated
numerically from the mean field equations ($\kappa=0.1,
\epsilon_J=0.05,\Gamma=1$).  } \label{fig:damping}
\end{figure}

The calculation of the damping proceeds as follows; the resonator
amplitude changes only slowly, so we can average the effect of the
systematic SSET motion over a single resonator period. Following the
calculation given in \cite{riha} we obtain an expression for the
amplitude-dependent frequency shift and damping introduced by the
SSET,
\begin{eqnarray}
\lefteqn{(\gamma_{SS}(z)+i\delta\omega_0(z))a= }\nonumber \\
&&i\frac{x_s\omega_0}{2x_q}\frac{(3\Gamma
+2i\omega_D)\epsilon_J^2}{(\Gamma+i\omega_D)^2}\sum\limits_{n}J_{-n}\left(z\right)J_{1-n}\left(z\right)\nonumber\\
&&\times\left(\frac{1}{\frac{\Gamma}{2}+i(\omega_Dn-\Delta)}
-\frac{1}{\frac{\Gamma}{2}+i(\omega_Dn+\Delta)} \right)\nonumber\\
\label{eq:damp}
\end{eqnarray}
In Fig. \ref{fig:damping}(a) we plot the damping as a function of
amplitude. The resonator $A$ amplitude is then found by solving the
self consistent equation for $a$ in the rotating frame,
\begin{eqnarray}
0&=&-i(\omega_0-\omega_D)a-
\gamma_{ex}a+iA_D\nonumber\\
&&-(\gamma_{SS}(z)+i\delta\omega_0(z))a.
\end{eqnarray}

With the resonator amplitude found, we can now go on to calculate
the cavity noise. Before doing this, we look at some simple
approximations to the damping. In the limit $z\to0$, we only need
include the $n=0,1$ terms in the sum and take the first order in $z$
to get a linear damping,
\begin{eqnarray}
\gamma_{SS}^{z\to0}&=&  \frac{-x_s^2\omega_0^2
2\Delta\epsilon_J^2\Gamma(13\frac{\Gamma^2}{4}+\omega_D^2+\Delta^2)}{x_q^2(\Delta^2\!+\!\frac{\Gamma^2}{4})
 (\Gamma^2\!+\!\omega_D^2)((\Delta^2\!+\!\frac{\Gamma^2}{4}\!-\!\omega_D^2)^2\!+\!\omega_D^2\Gamma^2)}\nonumber\\
 \label{eq:lindamp}
\end{eqnarray}
where we note that it is the driving frequency $\omega_D$ that
appears in the above expression rather than $\omega_0$ as the
driving field means the periodic motion is at this frequency. In the
absence of external driving, $\omega_D$ is replaced with $\omega_0$
in the above expression. Equation (\ref{eq:lindamp}) then reduces to
the previously known effective damping \cite{bia,SSET1,bc}, but in
the limit $\epsilon_J\ll \Gamma$ and extending to all resonator
frequencies \cite{cunpub} rather than just $\omega_0\ll\Gamma$. In
Fig. \ref{fig:damping}(b), the linear damping is compared to a value
obtained numerically from the mean-field equations for a range of
values of $\omega_D/\Gamma$.

We also find that we can get a relatively simple expression that is
exact to $\mathcal{O} (z^3)$, and valid for finite driving in the
limit $\omega_D \gg \Gamma$, and for a detuning that is not too
large, $|\Delta|\lesssim\omega_D$. Including only the $n=-1..2$
terms gives,
\begin{eqnarray}
\lefteqn{\gamma_{SS}(z)\approx }\nonumber \\
&&\gamma_{SS}^{z\to0} \frac{2J_0(z)J_1(z)}{z}
-\frac{4\Delta\epsilon_J^2x_s^2 J_1(z)J_2(z)}{x_q^2z}\times\cdots\nonumber\\
&&\Re\left\{ \frac{(3\Gamma
+2i\omega_D)3i\Gamma\omega_D}{(\Gamma\!+\!i\omega_D)\left((\frac{\Gamma}{2}\!-\!i\omega_D)^2\!+\!\Delta^2\right)\left((\frac{\Gamma}{2}\!+\!2i\omega_D)^2\!+\!\Delta^2\right)}
\right\} \nonumber\\\label{eq:nldampap}
\end{eqnarray}
This approximation to the amplitude-dependent damping is plotted in
Fig. \ref{fig:damping}(b).

The fluctuations in the charge act as an additional diffusion term
for the resonator field \cite{Blanter04}.
When there is no external driving, we
can insert the $z\to0,\omega_r\approx\omega_0$ limits of the damping
and the charge noise to obtain a simple effective temperature. We
find,
\begin{eqnarray}
(2\bar{n}_{SS}+1)&=&\frac{\frac{\Gamma^2}{4}+\Delta^2+\omega^2}{2\Delta\omega_0}
\end{eqnarray}
which again agrees with the previously known results
\cite{bia,SSET1,cunpub} for this system. Although this expression
has here been derived in the low-$\epsilon_J$ limit, other
calculations \cite{cunpub} suggest that this expression is exact for
all $\epsilon_J$.

 Interestingly, this expression for the effective temperature
is identical to the expression found when the mechanical resonator
is coupled to an optical cavity (or equivalent systems) rather than
an SSET \cite{Rae,MCCG,BlencoweBuks}, with the quasiparticle
tunneling rate $\Gamma$ simply replaced with the optical cavity
damping. This is a rather surprising result, as we have two very
different systems, an harmonic oscillator and a 3-level SSET,
providing the same effective temperature. The temperature is somehow
insensitive to the details of the measuring device - in particular
its finite level structure. This merits further investigation, and
it would be interesting to see if other related measuring devices
also lead to the same expression.

%

\section{Sensitive dependence of the sidebands on the driving amplitude}\label{sec:detect}

In this section we discuss in detail the rather sensitive dependence
of the height of the sideband peaks on the driving amplitude. Due to
the Bessel functions, the systematic and noisy response of the
charge at the sidebands oscillate rapidly as a function of $z$. We
investigate this and suggest that it could be used to detect the
presence of small changes in the amplitude of a large driving force.

The power of the systematic peaks in the frequency spectrum is
smaller than that in the noisy peaks by a factor of $\epsilon_J^2$,
and decreases as $1/z^2$, so we concentrate on the peaks in the
spectrum due to the fluctuating terms.

We envision driving the system at some (possibly large) value of
$z=z_0$. As the heights of the peaks at the sidebands oscillates
rapidly as a function of $z=z_0 +\delta z$, a small change of
$\delta z$ will lead to a large change in the height of the sideband
peaks.

The sensitivity of the detector depends on the height of the peak at
$z_0$ compared to the minimum height that the peak can have, i.e.
the floor in the frequency spectrum at this point. Equation
(\ref{eq:approxonpk}) gives an expression for the heights of the
peaks as a function of $z$, and thus we have a simple expression for
the ratio,
\begin{eqnarray}
R&\approx&\frac{4 J_n(z_{max})^2+2J_0(z_{max})^2}{4
J_n(z_{min})^2+2J_0(z_{min})^2} \label{eq:Rapprox1}
\end{eqnarray}
We find this gives a reasonable approximation for odd $n$, and for
low values of $z$ when $n$ is even. However, the approximation
breaks down when $z$ becomes large. When $z$ is large, the Bessel
functions tend to their asymptotic limit $J_n(z)\to \frac{2}{\pi
z}\cos(z-n\frac{\pi}{2}-\frac{\pi}{4})$, and we find that
$J_n^2=J_0^2$ for all even $n$, so Eq. (\ref{eq:Rapprox1}) diverges
at the point where $J_0(z_{min})=0$. In this limit, in order to
obtain the correct value for the ratio, we need to include terms to
the next order in $1/\omega_D$.

However, examining Eqs. (\ref{eq:noisesum2}) and
(\ref{eq:noisesum1}), we see that \emph{all} the terms in the sum
are of order $1/\omega_D^2$, {i.e.} to go to next order requires us
to perform an infinite sum. Fortunately, this can be done in the
large $z$ limit when the Bessel functions take their asymptotic
form.

For example, we can rewrite Eq. (\ref{eq:noisesum1}) at
$\omega\approx\omega_D n_e$, where $n_e$ is even and $n_e>0$,
separating out the odd and even terms to obtain, in the limit
$\Gamma,\Delta\ll\omega_D$,
\begin{eqnarray}
S_{cc}^{\eta_{02},\eta_{20}}(\omega)
&\approx&\frac{4}{(\omega_Dn_e)^2} \frac{J_{0}\left(z \right)^2
\Gamma\epsilon_J^2/(2\pi)}{\frac{\Gamma^2}{4}+ \Delta^2} \nonumber \\
&&+\frac{4}{(\omega_Dn_e)^2}\sum\limits_n \frac{J_{2n}\left(z
\right)^2 \Gamma\epsilon_J^2/(2\pi)}{( 2n\omega_D
)^2} \nonumber \\
&&+\frac{4}{(\omega_Dn_e)^2}\sum\limits_n \frac{J_{2n+1}\left(z
\right)^2 \Gamma\epsilon_J^2/(2\pi)}{( (2n+1)\omega_D )^2}.
\end{eqnarray}
We now use the fact that in the large $z$ limit, $J_n(z)^2$ has the
same value for all even and all odd $n$. This allows us to perform
the sums exactly, using expressions like
$\sum\frac{1}{m^2}=\frac{\pi^2}{6}$ to obtain,
\begin{eqnarray}
S_{cc}^{\eta_{02},\eta_{20}}(\omega) &\to&\frac{4J_{0}\left(z
\right)^2 \Gamma\epsilon_J^2}{(\omega_Dn_e)^2(2\pi)}
\left(\frac{1}{\frac{\Gamma^2}{4}+ \Delta^2} +\frac{\pi^2}{12\omega_D^2}\right) \nonumber \\
&&+\frac{4J_{1}\left(z \right)^2
\Gamma\epsilon_J^2}{(\omega_Dn_e)^2(2\pi)}
\left(\frac{\pi^2}{4\omega_D^2}\right) .
\end{eqnarray}
Performing similar calculations for Eqs. (\ref{eq:noisesum2}) and
(\ref{eq:noisesum3}) gives asymptotic forms for the charge noise for
at the odd, even and zeroth peak to leading order,
\begin{eqnarray}
 S_{cc}(0)&\to&\frac{\epsilon_J^2}{\pi z\Gamma
 }\frac{13\Gamma^2+4\Delta^2}{(\frac{\Gamma^2}{4}+\Delta^2)^2}\cos^2
 (z-\frac{\pi}{4})
+\frac{\pi\epsilon_J^2}{z\omega_D^2\Gamma}
\nonumber\\
 S_{cc}(n_e)&\to&\frac{\epsilon_J^2 \Gamma}{n_e^2\omega_D^2\pi z
}\frac{12}{\frac{\Gamma^2}{4}+\Delta^2}\cos^2
 (z-\frac{\pi}{4})
+\frac{3\pi\Gamma\epsilon_J^2}{z\omega_D^4n_e^2}
\nonumber\\
 S_{cc}(n_o)&\to&\frac{\epsilon_J^2\Gamma}{n_o^2\omega_D^2\pi z
 }\frac{4}{\frac{\Gamma^2}{4}+\Delta^2}(1+\sin^2
 (z-\frac{\pi}{4})).
\end{eqnarray}
The odd peaks have a ratio between maximum and minimum of order one,
so will not be of as much use. The even peaks have maxima at
$z=m\pi+\frac{\pi}{4}$, minima at $z=m\pi+\frac{3\pi}{4}$, and are
most sensitive to changes in $z$ at $z=\frac{m\pi}{2}$. The ratio
between the height of the peaks at their maxima and at their minima
takes a simple form when $n>0$. We find,
\begin{eqnarray}
R_{n_e>0}=\frac{\omega_D^2}{\frac{\Gamma^2}{4}+\Delta^2}
\end{eqnarray}
so the contrast between the peaks at their largest and smallest can
become large in the limit $\omega_D\gg\Gamma,\Delta$.

\section{Cavity Output Field} \label{sec:output}

In this section we describe how the noise spectrum of the field in
the cavity is transferred to the quantities that are actually
measured. In order to connect the dynamics of the resonator to
measured quantities we need to consider how the resonator is coupled
to the external world.  What form this takes depends on what the
resonator actually is. For example, if it is the microwave field in
a superconducting co planar cavity, we can use the input-output
formalism of quantum optics to relate the field in the microwave
cavity to the many mode fields in transmission lines connected to
the cavity. In this situation, the damping and noise on the
resonator is attributed to fields external to the cavity and these
external fields are ultimately what is measured. In the simplest
situation we could imagine a single side cavity with a quantum
limited input field. The output field from the cavity then contains
a component of the reflected input field as well as the field
transmitted from the cavity itself. This is the model we will adopt
here as we can easily apply the input-output theory of quantum
optics \cite {WM} as co-planar superconducting cavities are in the
highly underdamped limit appropriate for this formalism.

Another possible realization for the resonator is a nanomechanical
oscillator. In this case, we need an explicit transducer model for
the way in which the displacement of the nanomechanical resonator is
measured.  A typical example would be to capacitively couple the
nanomechanical resonator to a microwave cavity \cite{WoolleyMilb}.
In that case the nanomechanical resonator is coupled to more than
one bath: the finite-temperature mechanical bath in addition to its
irreversible coupling to the microwave field propagating into and
out of the transducer cavity.  For a fast and efficient measurement,
the microwave cavity would be strongly damped in which case the
resonator would see the bosonic bath due to the electromagnetic
fields on the transmission lines directly.

The quantum Langevin equation for the field is given in Eq.
(\ref{eq:fieldLGV}). In the microwave realization, we will assume a
single side cavity and that the only source of damping for the
cavity field is in fact its coupling to the input and output fields
at the open end of the cavity. In that case the noise operator,
$\eta_a$ is written in terms of the multi-mode field amplitude input
to the cavity: $\eta_a(t)=\sqrt{\gamma_{ex}}a_{in}(t)$. The output
field from the cavity is related to the input field and the
intracavity field by
\begin{equation}
a_o(t)=\sqrt{\gamma_{ex}}a(t)-a_{in}(t)
\end{equation}
In terms of the Fourier transformed operators, the input and output
fields are related by
\begin{eqnarray}
a_o(\omega)&=&-\frac{i\sqrt{\gamma_{ex}}
A_D\delta(\omega+\omega_D)}{\frac{\gamma_{ex}}{2}+i(\omega_D+\omega)}
+\frac{i\sqrt{\gamma_{ex}}\chi\sigma_{cc}(\omega)}{\frac{\gamma_{ex}}{2}+i(\omega_D+\omega)}\nonumber\\
&&+\frac{\frac{\gamma_{ex}}{2}-i(\omega_D+\omega)}{\frac{\gamma_{ex}}{2}+i(\omega_D+\omega)}a_{in}(\omega)
\label{output_amp}
\end{eqnarray}
where the last term represents a phase shift between incident and
reflected field components from a single-sided cavity, and
\begin{equation}
\chi=\frac{\omega x_s}{2x_q}
\end{equation}
and $\sigma_{cc}(\omega)$ is the Fourier transform of the island
charge operator $\sigma_{cc}(t)$. In general this is itself a
function of the intracavity field and so $\sigma_{cc}(\omega)$ is a
complicated convolution of a very nonlinear operator-valued function
of $a(\omega)$ and thus Eq.(\ref{output_amp}) is not an explicit
relation between the input and output field components. However if
we adopt the approximation implicit in Eq. (\ref{eq:resnoisenba}),
we can write it in terms of a systematic component and noise
component as

\begin{eqnarray}
a_o(\omega) & = & -\frac{i\sqrt{\gamma_{ex}}
A_D\delta(\omega+\omega_D)}{\frac{\gamma_{ex}}{2}+i(\omega_D+\omega)}
+\frac{i\sqrt{\gamma_{ex}}\chi\sigma^S_{cc}(\omega)}{\frac{\gamma_{ex}}{2}+i(\omega_D+\omega)}\nonumber\\
& &
+\frac{i\sqrt{\gamma_{ex}}\chi\sigma^\eta_{cc}(\omega)}{\frac{\gamma_{ex}}{2}+i(\omega_D+\omega)}
+\frac{\frac{\gamma_{ex}}{2}-i(\omega_D+\omega)}{\frac{\gamma_{ex}}{2}+i(\omega_D+\omega)}a_{in}(\omega)\nonumber\\
\end{eqnarray}
We thus see that the noise power spectrum for the field output from
the cavity is
\begin{eqnarray}
S_{out}(\omega)& = & \int_{-\infty}^\infty d\omega^\prime \langle a_o^\dagger(\omega),a_o(\omega^\prime)\rangle\nonumber\\
& = &
\bar{n}(\omega)+\frac{\chi^2(S_{c,c}^S(\omega)+S^\eta_{c,c}(\omega))}{(\omega-\omega_0)^2+\frac{\gamma_{ex}^2}{4}}
\end{eqnarray}
and we see that, for small enough $\bar{n}$, the noise of the cavity
is transferred to the output field, and thus the emitted field can
be used to detect the noise in the cavity.

\section{Conclusion} \label{sec:conclusions}

We have used a Langevin equation approach to investigate the
frequency spectrum of a superconducting single electron transistor
coupled to a resonator under periodic driving of the resonator. The
fluctuating noise terms allow us to describe the correlations in the
SSET due to the finite level structure. This approach allows the
calculation of the spectrum of the charge noise in the SSET and the
resulting effect on the resonator field in the limit of low
Josephson energy, $\epsilon_J\ll\Gamma$. We found that the charge
noise consists of a series of peaks at multiples of the driving
frequency, and calculated the heights at and between these noise
peaks
 in the limit of fast resonator
oscillation.

We have calculated the effect of the SSET on the cavity, and in
particular calculated an effective amplitude-dependent damping and
temperature that is valid for all resonator frequencies. We have
shown that the peaks at the sidebands depend rather sensitively on
the driving amplitude and show how this could provide a measure of
small changes in the driving amplitude.

We thank Andrew Armour and Thomas Harvey for helpful discussions.
D.A. Rodrigues acknowledges funding under EPSRC grant EP/D066417/1.

\appendix

\section{Derivation of the Langevin equations} \label{sec:apLV}

In this appendix we describe in more detail our use of the Langevin
equations and show how they can be derived from the master equation.
We show that this approach reproduces the results obtained by other
methods for an SSET uncoupled to a resonator.

We assume that the noise in the system is entirely determined by the
second order correlation functions for the dynamical variables. We
find that we can use Langevin equations as a useful tool for keeping
track of the dynamics of the first and second moments. If the
Langevin equations give the same equations of motion for the first
two moments as a master equation or Focker-Planck equation then, as
far as a noise calculation is concerned, the two are equivalent.
This equivalence is discussed in Ref. \cite{WM}, and a description
of a noise calculation given for the case when the conservative
terms and the diffusion in the Langevin equation are constant in
time. Here we focus on the situation where the external driving
means that these terms are time-dependent.

Although Langevin equations can be derived from a microscopic
picture, in which the noise terms $\eta(t)$ represent the
correlation functions of the external bath (in this case the
microscopic electron energy levels in the leads) here we take a
functional approach and consider them simply as a tool to allow us
to describe the correlation functions of the system. In this
approach we find that we can derive the properties of the noise
operators from the master equation. A general set of Langevin
equations $\bar{x}(t)$ with systematic evolution described by the
matrix $A(t)$ and fluctuations $\bar{E}(t)$ gives,
\begin{eqnarray}
\dot{\bar{x}}(t)&=&-A(t)\bar{x}+ \bar{E}(t) \label{eq:APL1}\\
\bar{x}(t)&=& \bar{x}(0)e^{-I(t)}+e^{-I(t)}\int\limits_0^t e^{I(t')}
\bar{E}(t')\rd t'\label{eq:APLint}
\end{eqnarray}
where $I(t)=\int\limits_0^t A(t') \rd t$. From Eq. (\ref{eq:APL1})
we can calculate equation of motion for the variance
$\chi(t)=\bar{x}(t)\bar{x}^T(t)$,
\begin{eqnarray}
\dot{\chi}(t)&=&\bar{x}(t)\left( -\bar{x}^T(t)A^T(t)+\bar{E}^T(t)
\right)\nonumber\\
&&+\left( -A(t)\bar{x}(t)+\bar{E}(t) \right)\bar{x}^T(t)
\end{eqnarray}
Inserting Eq. (\ref{eq:APLint}) and taking the ensemble average
gives,
\begin{eqnarray}
\av{\dot{\chi}(t)}&=&-(A(t)\av{\chi(t)}+\av{\chi(t)}A^T(t))\nonumber\\
&&+\left\langle \bar{E}(t)\left(\int\limits_0^t
\bar{E}^T(t')e^{I^T(t')}\rd
t'\right)e^{-I^T(t)}\right\rangle\nonumber\\
&&+\left\langle \left(e^{-I(t)}\int\limits_0^t e^{I(t')}
\bar{E}(t')\rd t'\right) \bar{E}^T(t)\right\rangle \label{eq:APvar2}
\end{eqnarray}
If the dynamics of the system is Markovian  (as in our master
equation), the fluctuating noise correlators will be
$\delta$-correlated. Indeed, a microscopic derivation shows that
$\delta$-correlated operators are obtained in exactly the limit that
the master equation becomes Markovian, {i.e.} when the decay of the
correlation functions of the leads is much more rapid than any
timescale in the system.

Writing the correlation matrix of the noise terms as
$\av{\bar{E}(t)\bar{E}^T(t')}=\delta(t-t')G(t)$, we obtain an
expression that relates the rate of change of the variance matrix to
the Langevin correlators,
\begin{eqnarray}
\av{\dot{\chi}(t)}&=&-(A(t)\av{\chi(t)}+\av{\chi(t)}A^T(t))+G(t)\label{eq:defG}
\end{eqnarray}
which reduces to the usual \cite{WM} $A\av{\chi}+\av{\chi}A^T=G$ for
the case of a time-independent $A$. The fluctuations represent
deviations from the mean field so we can also calculate $\mathbf G$
by comparing the true evolution of the second moments with the
mean-field-only evolution, which is sometimes more convenient in
practice.
 Defining
$\dot{\bar\varsigma}= -\mathbf{A}\bar{x}$, the elements of $\mathbf
G$ are also given by,
\begin{eqnarray}
{G_{ij}}= \dot{\av{x_i
x_j}}-\av{\dot{\varsigma_i}x_j}-\av{x_i\dot{\varsigma_j}}\label{eq:altG}.
\end{eqnarray}
Using either Eq. \ref{eq:defG} or Eq. \ref{eq:altG}, we find that
the correlators for the SSET are given by,
\begin{eqnarray}
G_{00,00} &=&-G_{00,11}=-G_{11,00}=\Gamma \av{\sigma_{11}(t)} \nonumber\\
G_{22,22} &=&-G_{11,22}=-G_{22,11}=\Gamma \av{\sigma_{22}(t) }\nonumber\\
G_{02,22} &=&-G_{02,11}=\Gamma \av{\sigma_{02}(t) }\nonumber\\
G_{22,20} &=&-G_{11,20}=\Gamma \av{\sigma_{20}(t) }\nonumber\\
G_{11,11} &=&\Gamma
(\av{\sigma_{11}(t)}+\av{\sigma_{22}(t)})\nonumber\\
G_{02,20} &=&\Gamma (\av{\sigma_{00}(t)}+\av{\sigma_{11}(t)}),
\label{eq:ssetcor}
\end{eqnarray}
with all other correlators equal to zero. We note that the
commutation relations are preserved, {e.g.} $G_{02,22}\neq
G_{22,02}$, retaining the quantum nature of the problem. It is also
worth re-emphasizing that this functional approach where we derive
the correlators through the equations of motion for the variance
means that we are only capturing 2nd-order correlations. This is
adequate for our purposes, as the noise only requires these terms,
but these Langevin equations give no information about higher-order
correlations.

We can check the validity of this approach to deriving the Langevin
correlators in a simple case where the noise can be calculated by
other methods. For the case of an undriven, uncoupled SSET,
inserting the Fourier transforms of Eqs. (\ref{eq:Lcs1}) to
(\ref{eq:L02}) along with the expressions for the correlators from
Eq. (\ref{eq:ssetcor}) in the expression for the noise, Eq.
(\ref{eq:FTdef1}), gives an expression that is mathematically
identical to the noise as calculated more directly using the
expression
\begin{eqnarray}
\av{\sigma_{ij}(t+\tau)\sigma_{kl}(t)}=\av{\exp(-A\tau)
\sigma_{ij}(t)\sigma_{kl}(t) }
\end{eqnarray}
which can be evaluated by exactly diagonalizing $A$. In particular,
this is true for all values of $\Gamma$ compared to the other SSET
timescales, $\Delta, \epsilon_J$. We can also use the converse of
this argument to confirm that the noise terms must indeed be
$\delta-$correlated.


\end{document}